\documentclass[12pt]{article}%
\usepackage{amsmath,latexsym}
\usepackage{graphicx}
\usepackage{amsmath}
\usepackage{amsfonts}
\usepackage{amssymb}%
\begin{document}

\title{{Neutron star interiors and topology change
}}
   \author{
Peter K.F. Kuhfittig*\\  \footnote{kuhfitti@msoe.edu}
 \small Department of Mathematics, Milwaukee School of
Engineering,\\
\small Milwaukee, Wisconsin 53202-3109, USA}

\date{}
 \maketitle
 \noindent

\begin{abstract}\noindent
Quark matter is believed to exist in the
center of neutron stars.  A combined
model consisting of quark matter and
ordinary matter is used to show that
the extreme conditions existing in the
center could result in a topology change,
that is, in the formation of wormholes.  
\end{abstract}

\section{Introduction}\label{S:introduction}
Wormholes are handles or tunnels in the
spacetime topology connecting different
parts of our Universe or different
universes.  The meticulous analysis in Ref.
\cite{MT88} has shown that such wormholes
may be actual physical objects permitting
two-way passage.

Also introduced in Ref. \cite{MT88} is the
metric for a static spherically symmetric
wormhole, namely
\begin{equation}\label{E:line1}
  ds^{2}=-e^{\phi(r)}dt^{2}+e^{\lambda(r)}dr^2
  +r^{2}(d\theta^{2}
  +\text{sin}^{2}\theta\,d\phi^{2}),
\end{equation}
where $e^{\lambda(r)}=1/(1-b(r)/r)$.  Here
$b=b(r)$ is the \emph{shape function} and
$\phi=\phi(r)$ is the \emph{redshift function},
which must be everywhere finite to prevent
an event horizon.  For the shape function
we must have $b(r_0)=r_0$, where $r=r_0$
is the radius of the \emph{throat} of the
wormhole; also, $b'(r_0)<1$ and $b(r)<r$
to satisfy the flare-out condition
\cite{MT88}. We are using geometrized
units where $G=c=1$.

While Morris and Thorne concentrated on
traversable wormholes suitable for humanoid
travelers (and possible construction by an
advanced civilization), a more general
question is the possible existence of
naturally occurring wormholes \cite{pK08}.
It has also been suggested that the
extreme conditions inside a massive
compact star may result in a topology
change \cite{DGL08}.  It is proposed in
this paper that a neutron star is a
candidate for such a topology change as
a result of \emph{quark matter}
\cite{nI70, HJ00, tS03, bM05}, which is
believed to exist at the center of neutron
stars \cite{PG10}.  To that end, we will
use a two-fluid model consisting of
ordinary matter and quark matter based
on the MIT bag model \cite{aG74}.  Two
cases will be considered: the
non-interacting and interacting two-fluid
models.  (We will concentrate mainly on
the former case.)  The reason for having
two cases is that quark matter is a viable
candidate for dark matter \cite{aB99}, \
albeit under different circumstances,
and so is likely to be weakly interacting.

A combined model consisting of neutron-star
matter and of a phantom/ghost scalar field
yielding a nontrivial topology is discussed
in Ref. \cite{vD12}.  The model is referred
to as a neutron-star-plus-wormhole
system.

\section{Basic equations}
For our basic equations we follow Ref.
\cite{fR12}.  The energy momentum tensor
of the two-fluid model is given by
\begin{equation}\label{E:rho}
   T^0_0\equiv\rho_{\text{effective}}
       =\rho+\rho_q,
\end{equation}
\begin{equation}\label{E:p}
   T^1_1=T^2_2\equiv-p_{\text{effective}}
       =-(p+p_q).
\end{equation}
In Eqs. (\ref{E:rho}) and (\ref{E:p}),
$\rho$ and $p$ correspond to the respective
energy density and pressure of the baryonic
matter, while $\rho_q$ and $p_q$ correspond
to the energy density and pressure of the
quark matter.  The left-hand sides are the
effective energy density and pressure,
respectively, of the combination.

The Einstein field equations are listed
next \cite{fR12}:

\begin{equation}\label{E:Einstein1}
 e^{-\lambda}
 \left(\frac{\lambda^\prime}{r} - \frac{1}{r^2}
 \right)+\frac{1}{r^2}=8\pi (\rho +\rho_q),
\end{equation}
\begin{equation}\label{E:Einstein2}
  e^{-\lambda}
  \left(\frac{\phi'}{r}+\frac{1}{r^2}\right)
  -\frac{1}{r^2}=8\pi(p+p_q),
\end{equation}
and
\begin{equation}\label{E:Einstein3}
  \frac{e^{-\lambda}}{2}
  \left[\frac{(\phi')^2-\lambda'\phi'}{2}
  +\frac{\phi'-\lambda'}{r}+\phi''
  \right]=8\pi (p+p_q),
\end{equation}
recalling that $T^1_1=T^2_2$.

In the MIT bag model, the matter equation
of state is given by
\begin{equation}\label{E:bag}
  p_q=\frac{1}{3}(\rho_q-4B),
\end{equation}
where $B$ is the bag constant, which we take
to be $145\,\text{MeV}/(\text{fm})^3$
\cite{aG74, fR11}.  Since quarks are part
of the standard model, we assume that
$p_q>0$; hence $\rho_q>4B$.  For normal
matter we use the somewhat idealized
equation of state \cite{fR09}
\begin{equation}\label{E:EoS}
   p=m\rho, \quad 0<m<1.
\end{equation}

Since we are assuming the pressure to
be isotropic, the conservation equation
is \cite {fR12}
\begin{equation}\label{E:conservation}
  \frac{d(p_{\text{effective}})}{dr}
  +\frac{1}{2}\phi'(\rho_{\text{effective}}
  +p_{\text{effective}})=0.
\end{equation}

\section{Solutions of basic equations}

As noted in the Introduction, we will
concentrate mainly on the non-interacting
fluid model.  This means that the two fluids,
consisting of normal matter and quark
matter, do not interact.  The resulting
conservation equations are therefore
independent of each other.  Using Eqs.
(\ref{E:bag}) and (\ref{E:EoS}), we have
\begin{equation}
  \frac{d\rho}{dr}+\phi'\left(
  \frac{1+m}{2m}\right)\rho=0
\end{equation}
and
\begin{equation}
   \frac{d\rho_q}{dr}+2\phi'(\rho_q-B)=0.
\end{equation}
The solutions are
\begin{equation}\label{E:rho1}
  \rho=\rho_0e^{-\phi(1+m)/2m}
\end{equation}
and
\begin{equation}\label{E:rho2}
    \rho_q=B+\rho_{(q,0)}e^{-2\phi},
\end{equation}
where $\rho_0$ and $\rho_{(q,0)}$ are
integration constants.  Equations
(\ref{E:Einstein1}), (\ref{E:Einstein2})
anf (\ref{E:Einstein3}) now give
\begin{equation}\label{E:linear}
   e^{-\lambda}\left[-\frac{\lambda'\phi'}{2}
   +\frac{(\phi')^2}{2}+\phi''+
   \frac{2\phi'}{r}\right]=8\pi[\rho(1+3m)+
   (2\rho_q-4B)].
\end{equation}
It now becomes apparent that this equation
is linear in $e^{-\lambda}$ once it is
written in the following form:
\begin{equation*}
   (e^{-\lambda})'+e^{-\lambda}\left(\phi'
   +\frac{2\phi''}{\phi'}+\frac{4}{r}\right)
   =\frac{2}{\phi'}(8\pi)[\rho(1+3m)
   +(2\rho_q-4B)].
\end{equation*}
The integrating factor is
\begin{equation*}
   \text{I.F.}=e^{\phi+2\,\text{ln}\,\phi'
   +4\, \text{ln}\,r}=e^{\phi}(\phi')^2r^4,
\end{equation*}
leading at once to
\begin{equation*}
   \frac{d}{dr}[e^{-\lambda}e^{\phi}(\phi')^2
   r^4]=e^{\phi}(\phi')^2r^4
   \left(\frac{2}{\phi'}\right)(8\pi)
   [\rho(1+3m)+(2\rho_q-4B)].
\end{equation*}
To write the solution for $e^{-\lambda}$, we
first define $F(r)=\rho(1+3m)+(2\rho_q-4B)$,
which becomes, after substituting the
solutions $\rho$ and $\rho_q$,
\begin{equation*}
   F(r)=\rho_0e^{-\phi(1+m)/2m}(1+3m)
   +2\rho_{(q,0)}e^{-2\phi}-2B.
\end{equation*}
Now we have
\begin{equation}
   e^{-\lambda}=\frac{16\pi}{e^{\phi}
   (\phi')^2r^4}\int^r_{r_0}e^{\phi}
   \phi'r_1^4F(r_1)dr_1.
\end{equation}
The lower limit $r=r_0$ is the radius
of the throat of the wormhole, as we
will see.

As noted in Sec. \ref{S:introduction},
the shape function $b=b(r)$ is obtained
from $e^{-\lambda(r)}$, so that
\begin{equation}
   b(r)=r(1-e^{-\lambda(r)}).
\end{equation}
We obtain
\begin{equation}
   b(r)=r\left[1-\frac{16\pi}{e^{\phi}
   (\phi')^2r^4}\int^r_{r_0}
   e^{\phi}\phi'r_1^4F(r_1)dr_1\right].
\end{equation}
Note especially that $b(r_0)=r_0$, which
characterizes the throat of the wormhole.

To study the other requirement,
$0<b'(r_0)<1$, we start with
\begin{equation*}\label{E:bprime}
   b'(r)=1-e^{-\lambda(r)}+r\left[
   -\frac{d}{dr}e^{-\lambda(r)}
   \right],
\end{equation*}
which leads to
\begin{equation}\label{E:bprime}
   b'(r_0)=1+r_0\frac{16\pi}{\phi'(r_0)}
   \left[-\rho_0
   e^{-\phi(r_0)(1+m)/2m}(1+3m)
   -2\rho_{(q,0)}e^{-2\phi(r_0)}
   +2B\right].
\end{equation}
Given that $\phi'$, the gradient of
the redshift function, is related to
the tidal force, one would expect
its magnitude to be extremely large.
So $b'(r_0)$ will indeed be less
than unity if the expression inside
the brackets in Eq. (\ref{E:bprime})
is negative and sufficiently small
in absolute value.  If this is the
case, then the flare-out condition
is met, that is,
\begin{equation}
   \frac{r_0b'(r_0)-b(r_0)}
   {2[b(r_0)]^2}<0.
\end{equation}
Since this indicates a violation
of the null energy condition, the
flare-out condition is the primary
prerequisite for the existence of
wormholes \cite{MT88}.  (The
condition $b(r)<r$ will be checked
later.)

 Unfortunately, the constants $B$
 and $\rho_{(q,0)}$ are also large,
 so that Eq. (\ref{E:bprime}) must
 be examined more closely.

 \section{The gradient of the redshift
 function}

To analyze Eq. (\ref{E:bprime}) we start
with the gradient of $\phi$ \cite{WNR07}:
\begin{equation*}
   \frac{d\phi}{dr}=\left[\frac{2Gm(r)}{c^2r^2}
   +8\pi r\frac{G}{c^4}p\right]
   \frac{1}{1-2Gm(r)/c^2r},
\end{equation*}
where $m(r)$ is the total mass energy
inside the radial distance.  (This
formula is also used in the derivation
of the Tolman-Oppenheimer-Volkoff
equation.)  In geometrized units we
get for $r=r_0$
\begin{equation}\label{E:gradient}
   \phi'(r_0)=
   \frac{2m(r_0)/r_0^2+8\pi r_0p(r_0)}
  {1-2m(r_0)/r_0}.
\end{equation}

Since we are now concerned with what
is going to become the throat $r=r_0$,
we need to evaluate $\rho_0$ and
$\rho_{(q,0)}$ in Eqs. (\ref{E:rho1})
and (\ref{E:rho2}) to obtain
\begin{equation}
   \rho_0=\rho(r_0)
   e^{\phi(r_0)(1+m)/2m}
\end{equation}
and
\begin{equation}
   \rho_{(q,0)}=e^{2\phi(r_0)}
   [\rho_q(r_0)-B].
\end{equation}
So Eq. (\ref{E:bprime}) becomes
\begin{equation}
   b'(r_0)=1+
   \frac{16\pi r_0[-\rho(r_0)(1+3m)
        -2\rho_q(r_0)+4B]}
   {[2m(r_0)/r_0^2+8\pi r_0p(r_0)]
         /[1-2m(r_0)/r_0]}.
\end{equation}
We already know that $p=m\rho$
and that $\rho_q>4B$, so that
the numerator is indeed negative.
For the purpose of comparison, we
will simply assume for now that
$\rho_q(r_0)=4B$.  (See Remark 1
below.)  Then
\begin{equation}\label{E:flare}
   b'(r_0)=1+
   \frac{16\pi[-\rho(r_0)(1+3m)-4B]}
   {[2m(r_0)/r_0^3+8\pi p(r_0)]
        /[1-2m(r_0)/r_0]}.
\end{equation}
Given that $\frac{1}{2}b(r)$ is
the effective mass inside a sphere
of radius $r$, $1-2m(r_0)/r_0=0$.
So near $r=r_0$, the fraction in
Eq. (\ref{E:flare}) is small and
negative.  Observe also that 
according to Eq. (\ref{E:gradient}), 
near $r=r_0$, $\phi'$ is indeed 
large, as expected.

We conclude that $b'(r_0)<1$.  So
in the neighborhood of the throat,
$b(r)/r<1$.  To show that this
inequality holds away from the
throat, observe that
\[
 \frac{1}{2}b(r)= \frac{1}{2}b(r_0)
  +\int^r_{r_0}4\pi r_1^2\rho(r_1)dr_1
\]
and
\[
   \frac{1}{2}b'(r)=4\pi r^2\rho(r).
\]
Since $\rho(r)$ decreases as $r$
increases, $b'(r)$ is positive, but
decreasing, so that $b(r)/r$ remains
less than unity for $r>r_0$.

\emph{Remark 1:} Given that the
density of nuclear matter ranges
from $6\times10^{17}$\,$\text{kg/m}^3$
to $8\times10^{17}$\,$\text{kg/m}^3$,
we may take $\rho_q$ to be about
$10^{18}$\,$\text{kg/m}^3$.  So
 $\rho_q(r_0)$
is quite close to $4B$, thereby
justifying the assumption
$\rho_q(r_0)\approx 4B$.

\section{Completing the wormhole
   structure}
Since $r=r_0$ is the throat of the
wormhole, it follows from the
definition of throat that the
interior $r<r_0$ is outside the
manifold.  Although not part of
the wormhole spacetime, it still
contributes to the gravitational
field.  This can be compared to
a thin-shell wormhole from a
Schwarzschild black hole
\cite{mV89}: while not part of
the manifold, the black hole
generates the gravitational
field.

Since the throat of the wormhole
is deep inside the neutron star,
it cannot be directly observed.
According to Ref. \cite{vD12},
however, there may still be
observable tell-tale signs: if
two neutron stars are connected
by a wormhole, they would have
identical, or nearly identical,
characteristics.  An even stronger
indication would be any observed
variation in the mass of a neutron
star.

\emph{Remark 2:}  The quark-matter
core, being surrounded by nuclear
matter, raises a question regarding
the interface region.  As noted in
the Introduction, quark matter is
believed to be weakly interacting.
So the most likely result is a
drop in the energy density in the
neighborhood of the core's surface.

We conclude with some brief comments
on the interacting case.  If the two
fluids are assumed to interact, then
the conservation equations take on
the following forms \cite{fR12, PW09}:
\begin{equation}
   \frac{d\rho}{dr}+\phi'\left(
   \frac{1+m}{2m}\right)\rho=Q
\end{equation}
and
\begin{equation}
   \frac{d\rho_q}{dr}+2\phi'
   (\rho_q-B)=-3Q.
\end{equation}
The quantity $Q$ expresses the interaction
between the two types of matter and falls
off rapidly as $r$ increases.
Furthermore, since the interaction is
assumed to be relatively weak, $Q$ is a
very small quantity compared to $B$.  So
the presence of the quantity $Q$ is going
to have little effect.

\section{Conclusion}
Quark matter is believed to exist in
the center of neutron stars.  The analysis
in this paper is therefore based on a
two-fluid model comprising ordinary and
quark matter with an isotropic matter
distribution.  It is shown that the
extreme conditions existing in the
center could result in a topology
change, that is, a neutron star with
a quark-matter core could give rise
to a wormhole.

\end{document}